\documentclass[prb,twocolumn,superscriptaddress]{revtex4}

\usepackage{graphicx}
\usepackage{bm}

\begin{document}
\newcommand{\be}{\begin{equation}}
\newcommand{\ee}{\end{equation}}
\newcommand{\bea}{\begin{eqnarray}}
\newcommand{\eea}{\end{eqnarray}}


\title{Optical cooling of atoms in microtraps by time-delayed reflection}

\author{Peter Horak}
\affiliation{Optoelectronics Research Centre, University of
Southampton, Southampton SO17 1BJ, United Kingdom}

\author{Andr\'e Xuereb}
\affiliation{School of Physics and Astronomy, University of
Southampton, Southampton SO17 1BJ, United Kingdom}

\author{Tim Freegarde}
\affiliation{School of Physics and Astronomy, University of
Southampton, Southampton SO17 1BJ, United Kingdom}


\date{\today}

\begin{abstract}
We present a theoretical analysis of a novel scheme for optical
cooling of particles that does not in principle require a closed
optical transition. A tightly confined laser beam interacting with a
trapped particle experiences a phase shift, which upon reflection
from a mirror or resonant microstructure produces a time-delayed
optical potential for the particle. This leads to a nonconservative
force and friction. A quantum model of the system is presented and
analyzed in the semiclassical limit.

\vspace{5mm} \noindent

\end{abstract}

\maketitle


\section{Introduction}

Standard techniques for optical cooling of atoms mostly rely on
spontaneous emission from the atom to carry away the excess
momentum. This basic process has been demonstrated to be highly
efficient for cooling of two-level or multi-level atoms and for
various laser configurations in one, two, or three dimensions,
applying various polarization states and frequencies \cite{Nobel}.
However, a common requirement is for the atom to exhibit a single,
albeit possibly degenerate, ground state such that no atomic
population is lost from the cooling cycle by population transfer
into internal states decoupled from the laser light. These cooling
mechanisms are thus not well suited for most atomic species and
hardly for any molecules at all.

Cavity-mediated cooling mechanisms \cite{PRL,Vuletic,JOSAB,Maunz}
have been suggested to address these shortcomings of laser cooling
methods. In this case, only a dipole interaction between the
particles and a near resonant cavity mode is required. While this
can be fulfilled for a much larger variety of particles, resonator
alignment and loading of the particles into the small mode volume of
an appropriate cavity is difficult.

We have recently proposed an alternative method for cooling
particles that exhibit an electric dipole moment but no closed
transitions \cite{mirror-cooling}. In this `mirror-mediated cooling'
scheme, a laser-driven particle interacts with its own image in a
mirror and the time-delay incurred during the reflection is
exploited to introduce a non-conservative element into the dipole
force, thereby leading to friction and cooling. This time delay may
be induced by a delay line, e.g., an optical fiber, but more
conveniently could arise from an integrated optical micro-resonator
\cite{Vahala,Schmiedmayer,Reichel,Mabuchi,Baumberg}, possibly
combined with a plasmonic microstructure for local field enhancement
\cite{Alu,Fischer}.

Here we investigate the challenges involved in modeling this novel
cooling mechanism. We discuss how the required processes differ from
free-space or cavity-mediated laser-cooling methods and propose a
theory to overcome these new difficulties. We finally discuss the
main results obtained from our method.


\section{Conceptual differences between optical cooling methods}

\begin{figure}[bth]
  \includegraphics[width=6.5cm]{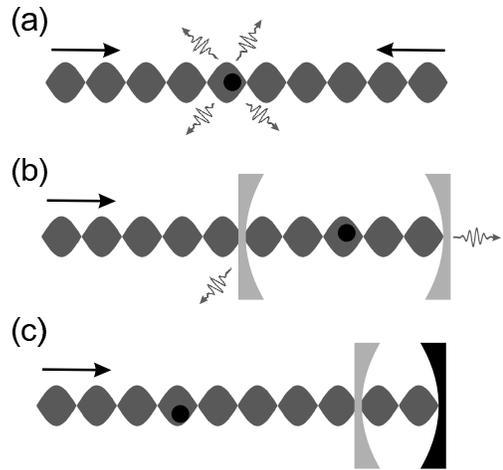}
  \caption{Schematics of optical cooling methods: (a) Free-space
  laser cooling, (b) cavity-mediated cooling of a particle between
  two partially reflective mirrors, (c) mirror-mediated cooling of
  a particle using one perfectly reflecting and one partially
  reflective mirror.}
  \label{fig:schematics}
\end{figure}

We start our discussion with a brief review of the fundamental
concepts behind three different laser cooling schemes: (i)
free-space laser cooling, (ii) cavity-mediated cooling, and (iii)
mirror-mediated cooling.

Free-space laser cooling methods, such as Doppler cooling
\cite{Gordon}, polarization gradient cooling \cite{polgrad}, or
velocity-selected coherent population trapping \cite{vscpt}, utilize
a number of laser beams focused from various directions on a small
sample of atoms, as illustrated in Fig.\ \ref{fig:schematics}(a). In
each case, the laser and atomic configuration is set up in such a
way that the atom-light coupling is velocity-dependent. For example,
the photon scattering rate or the spatial positions where scattering
predominantly happens can depend on the atomic velocity, leading to
an average net cooling effect. The excess momentum of hot atoms is
thereby transferred to spontaneously emitted photons. A mathematical
description of the cooling mechanism thus contains a number of
classical, stationary laser beams, and a quantum description of the
atomic degrees of freedom. To remove the excess momentum, the
excited states of the atoms are coupled to a heat bath, which is
assumed to be memory-less (Markovian) \cite{Gardiner}.

In cavity-mediated cooling schemes \cite{PRL,Vuletic,JOSAB,Maunz},
Fig.\ \ref{fig:schematics}(b), the atoms sit inside an optical
resonator and are coupled to one or a few resonator modes, while
laser beams are used to pump the resonator through its mirrors.
Dipole coupling transfers excess momentum coherently from the atoms
to the cavity modes, which subsequently decay through the
partially-transmitting mirrors. Mathematically, the classical
stationary laser beams are now coupled to a quantum system
comprising the atom degrees of freedom and the quantum state of the
discrete cavity modes. The modes in turn are coupled to a Markovian
heat bath.

In mirror-mediated cooling, finally, the particles sit outside a
coherent delay device, Fig.\ \ref{fig:schematics}(c), which could be
as simple as a single mirror, a piece of optical fiber with an
inscribed Bragg grating, or formed by an integrated optical
resonator on a chip
\cite{Vahala,Schmiedmayer,Reichel,Mabuchi,Baumberg}. Via the dipole
coupling, the particle imprints a phase on the pump beam which
returns to the position of the particle after reflection with a time
delay. Hence, the total dipole potential experienced by the particle
depends on the interference of the incoming beam with the
time-delayed reflected beam. If the particle has moved during the
light round trip, this leads to a non-conservative force that can be
exploited for extracting energy from the particle motion. A
mathematical description of this situation therefore requires two
key ingredients, different from free-space cooling and
cavity-mediated cooling: (i) the pump beam itself must be described
as dynamic, and (ii) cooling relies on the system state at an
earlier time, thus it is non-Markovian. On the other hand, no heat
bath is required.

Instead of introducing a system memory in time, we may also
decompose the field into a continuum of modes. The model presented
here will follow this latter approach. For the sake of simplicity,
we model the polarizable particle as a single two-level atom and
assume that the time delay arises from a significant distance
(several meters) between the atom and the mirror. However, it is
envisaged that in a practical realization the time delay will arise
from reflection by an integrated micro-optical resonator or a
similar structure. Moreover, we restrict the analysis to a single
spatial dimension. The electromagnetic modes at angular frequency
$\omega$ thus are standing waves with mode functions
$f(\omega,x)=\sin(\omega x/c)$ if the mirror is assumed at $x=0$.
Upon adiabatic elimination of the internal degrees of freedom of the
atom, and treating atomic motion semiclassically, we obtain a
continuum quantum model governed by the Hamiltonian
 \bea
 \hat{H} &=&
 \int\hbar(\omega-\omega_0)\hat{a}^{\dagger}(\omega)\hat{a}(\omega) \mathrm{d}\omega
 \nonumber \\
 & & +\hbar \frac{g^2}{\Delta}\int\!\!\!\int\sin\frac{\omega_1 x}{c}\sin\frac{\omega_2 x}{c}
     \hat{a}^{\dagger}(\omega_1)\hat{a}(\omega_2) \mathrm{d}\omega_1
     \mathrm{d}\omega_2 \quad\quad
 \label{eq:hamiltonian}
 \eea
where $\hat{a}(\omega)$ and $\hat{a}^{\dagger}(\omega)$ are the mode
annihilation and creation operators, $\Delta$ is the detuning of the
atom from the driving laser, and $g$ is the atom-field coupling
constant. For a two-level atom with transition wavelength $\lambda$
and excited state decay rate $2\Gamma$, this coupling constant is
related to the mode beam waist $w$ by \cite{scat1d}
 \be
 2\pi g^2 = \Gamma \frac{4\sigma_a}{\pi w^2}
 \label{eq:g}
 \ee
where $\sigma_a=3\lambda^2/(2\pi)$ is the atomic radiative cross
section. The force operator describing the action of the field on an
atom at position $x$ is derived from the Hamiltonian as
 \be
 \hat{F}(x) = -\frac{\mathrm{d}\hat{H}}{\mathrm{d} x}.
 \label{eq:force}
 \ee

We note that the general form of the inter-modal coupling given by
the Hamiltonian (\ref{eq:hamiltonian}) is valid for any point-like
dipole scatterer, and thus will also hold for general multi-level
atoms and molecules. Moreover, in order to calculate the force
(\ref{eq:force}) we are interested only in the time evolution of the
field in the vicinity of the particle. Hence, it is mainly the
relative phase between the mode functions at this position which is
important. We thus conclude that, instead of free-space propagation,
any dispersive optical device could be used as a delay element,
which in particular includes resonant micro- or nanostructures.


\section{Perturbative solution of the model}

In the following we apply perturbation theory to the model
introduced above to derive the basic properties of the proposed
cooling method. We work in the Heisenberg picture where the mode
operators become time dependent with the dynamics governed by
 \bea
 \frac{\mathrm{d}}{\mathrm{d}t} \hat{a}(\omega,t) &=& \frac{i}{\hbar}\left[\hat{H}, \hat{a}(\omega,t)\right]
 \nonumber \\
 & = & -i(\omega-\omega_0)\hat{a}(\omega,t)\nonumber\\
 & & -\ i \frac{g^2}{\Delta}\sin\frac{\omega x}{c}\int\sin\frac{\omega_1 x}{c}
     \hat{a}(\omega_1,t) \mathrm{d}\omega_1.
 \label{eq:dadt}
 \eea
Next, we apply a semiclassical approximation for the field modes
assuming that the state of every mode is given by a coherent state
at all times. This is tantamount to replacing the operator
$\hat{a}(\omega,t)$ by its expectation value $a(\omega,t)$. We
assume that the atom follows a linear trajectory $x(t)=x+v(t-t_0)$
such that at a time $t_0\gg 2\tau$ the atom is at position $x$.
Here, $2\tau=2x/c$ is the round-trip time of the delayed reflection.
We then expand the mode amplitudes into powers of both the coupling
$g^2/\Delta$ and the atom velocity $v$,
 \be
 a(\omega,t) = a_0(\omega,t)
    + \frac{g^2}{\Delta} \left[a_1(\omega,t) + v
    b_1(\omega,t)\right] + \cdots .
 \label{eq:ansatz}
 \ee
The zeroth order term in $g^2/\Delta$ corresponds to the field
without back action of the atom. It is thus independent of $v$ and
represents the unperturbed driving laser field, which is assumed to
be monochromatic,
 \be
 a_0(\omega,t) = A\delta(\omega-\omega_0).
 \label{eq:a0}
 \ee
Here, $|A|^2$ gives the pump power in units of photons per second.
Inserting (\ref{eq:ansatz}) with (\ref{eq:a0}) into (\ref{eq:dadt}),
one can derive the first order terms in $g^2/\Delta$ analytically by
perturbation theory yielding
 \bea
 a_1(\omega,t_0) &=& A\frac{\exp\left[-i(\omega-\omega_0)t_0\right]-1}{\omega-\omega_0}
    \sin\frac{\omega x}{c}\sin\frac{\omega_0 x}{c},\nonumber\\
 & &  \label{eq:a1} \\
 b_1(\omega,t_0) &=& \frac{A}{c}\left[
    \omega_0\sin\frac{\omega x}{c}\cos\frac{\omega_0 x}{c}
    +\omega\cos\frac{\omega x}{c}\sin\frac{\omega_0 x}{c}
    \right]\nonumber \\
 & & \times\frac{
        -1 +[1+i(\omega-\omega_0)t_0]\exp[-i(\omega-\omega_0)t_0]
     }{(\omega-\omega_0)^2}.\nonumber\\
 & & \label{eq:b1}
 \eea
Expressions (\ref{eq:ansatz})-(\ref{eq:b1}) can then be inserted
into (\ref{eq:force}) to calculate the leading terms of the force
experienced by the atom at position $x$ at time $t_0$. Most
interesting is the first-order term of the force in velocity $v$,
since this gives the linear friction force and thus describes
heating or cooling of the atom in the proposed setup. We find
 \be
 F_v(x,t_0) = 2\pi\hbar k_0^2v\tau|A|^2\frac{g^4}{\Delta^2}\sin(4k_0x)
 \ee
in leading order of $x/\lambda$. Here, $k_0=\omega_0/c$ is the pump
wavenumber. Note that $F_v$ is independent of time. We now define
the spatially dependent friction coefficient $\rho(x)$ by the
relation $F_v(x)=-\rho(x) mv$ and introduce the atomic saturation
parameter $s=|A|^2g^2/\Delta^2$ at the maximum of the standing-wave
pump. Together with (\ref{eq:g}) this yields
 \be
 \rho(x) = -4s\Gamma\frac{\sigma_{\text{a}}}{\pi w^2}
           \frac{\hbar k_0^2}{m}\tau\sin(4k_0x)
 \label{eq:rhox}
 \ee
which has units of s$^{-1}$ and therefore relates to the inverse of
the cooling or heating time.

\begin{figure}[tbh]
  \includegraphics[width=7cm]{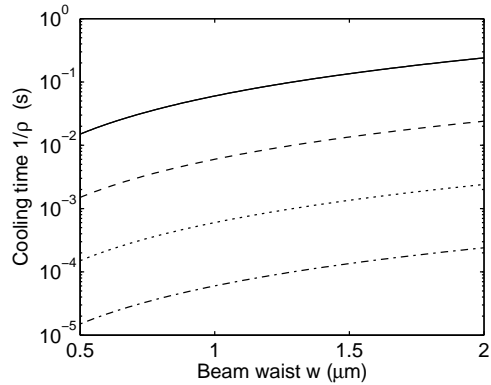}
  \caption{Cooling time of rubidium atoms at position of maximum
  friction versus beam waist $w$ for $s=0.1$ and $\tau=1$ ns (solid
  line), 10 ns (dashed), 100 ns (dotted), and 1 $\mu$s (dash-dotted).}
  \label{fig:coolingtime}
\end{figure}

Fig.\ \ref{fig:coolingtime} shows the dependence of the 1/e cooling
time for rubidium atoms at a point of maximum friction,
$\sin(4k_0x)=-1$, on the beam waist and for different delay times
ranging from 1 ns to 1 $\mu$s. Note that this plot assumes constant
saturation $s$ and thus the pump laser power is assumed to increase
linearly with the mode area. The figure predicts that cooling times
of the order of ms can be achieved if the pump is tightly focussed
at the atom and the delay time is of the order of tens of ns. Longer
delays lead to faster cooling. These conditions are comparable with
those explored experimentally in, for example, Ref.\
\onlinecite{Eschner}.

\begin{figure}[tbh]
  \includegraphics[width=7cm]{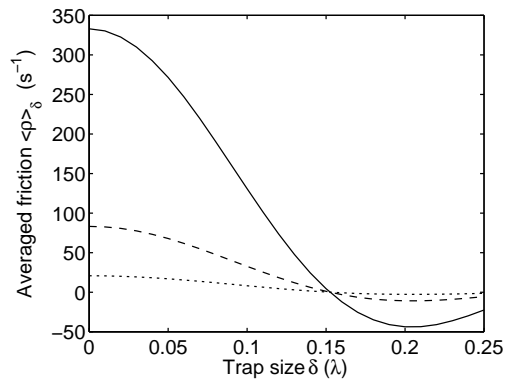}
  \caption{Averaged friction coefficient
  $\langle\rho(x)\rangle_{\delta}$ for a trap centered at a point of
  maximum friction versus trap size $\delta$ for $s=0.1$, $\tau=10$
  ns, and $w=0.5$ $\mu$m (solid line), $w=1$ $\mu$m (dashes), $w=2$ $\mu$m
  (dotted).}
  \label{fig:frictionav}
\end{figure}

An important feature of the friction coefficient (\ref{eq:rhox}) is
its spatial dependence with $\sin(4k_0x)$. This implies that the net
friction for an extended spatial distribution is zero. Significant
cooling via this method thus requires localizing the atom in an
additional trap. In practice this could be achieved by an additional
far-off resonant beam either propagating parallel to the mirror or
forming another standing wave superimposed on the driving beam for
the cooling. Alternatively, on-chip microtraps can be utilized in
conjunction with integrated time-delay reflectors. In the following,
we will assume a harmonic trapping potential. We can then calculate
an averaged friction coefficient, defined via the loss of kinetic
energy over one oscillation, as
 \bea
 \langle\rho(x)\rangle_{\delta} &=& -4s\Gamma\frac{\sigma_{\text{a}}}{\pi w^2}
           \frac{\hbar k_0^2}{m}\tau \frac{1}{2\pi} \nonumber \\
 & & \times\int_0^{2\pi}\sin[4k_0x+4\delta\sin(T)]\cos^2(T)
 \mathrm{d}T\,\,\,\quad
 \label{eq:rhoav}
 \eea
where $\delta$ is the maximum displacement of the atom from the trap
center during an oscillation. Fig.\ \ref{fig:frictionav} shows the
averaged friction coefficient versus trap size, assuming that the
trap center is at a point of maximum friction. As expected, for
increasing $\delta$ the cooling force decreases since the atom moves
away from the point of optimum cooling. For trap sizes
$\delta\gtrsim 0.15\lambda$ cooling finally turns into heating as
the atom spends more time in regions of the standing-wave pump where
the mirror-mediated force accelerates the atom.

\begin{figure}[tb]
  \includegraphics[width=7cm]{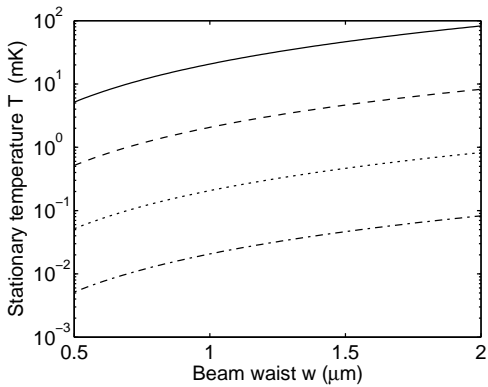}
  \caption{Stationary temperature T at position of maximum
  friction versus beam waist $w$ for $s=0.1$ and $\tau=1$ ns (solid
  line), 10 ns (dashed), 100 ns (dotted), and 1 $\mu$s (dash-dotted).}
  \label{fig:temperature}
\end{figure}

So far, we have neglected momentum diffusion due to spontaneous
scattering of photons by the atom, which together with the friction
coefficient determines the stationary temperature achievable with
this system. To zeroth order in the atom-field coupling, momentum
diffusion is due to the interaction with the unperturbed pump field.
We may thus expect diffusion to be identical to that observed in
free-space Doppler cooling, where the diffusion constant is given by
\cite{Gordon}
 \be
 D = \hbar^2k_0^2\Gamma s.
 \ee
From this the stationary temperature is obtained as
 \be
 k_BT = \frac{D}{m\rho(x)} = \frac{\hbar}{\tau}
 \frac{\pi w^2}{4\sigma_a}\frac{-1}{\sin(4k_0x)}
 \label{eq:Tanal}
 \ee
where for simplicity we have used the non-averaged value of the
friction (\ref{eq:rhox}). This is a remarkably simple expression
which, apart from the spatial dependence, only depends on the delay
time and the beam cross section. Fig.\ \ref{fig:temperature} shows
the stationary temperatures corresponding to the friction curves of
Fig.\ \ref{fig:coolingtime}. For realistic parameters these simple
analytic results predict stationary temperatures of the order of mK
or even slightly below and thus less than an order of magnitude
above the Doppler limit of 141 $\mu$K for Rb atoms. However, we
emphasize again that the cooling in our scheme is based on the
dipole force, in contrast to free-space Doppler cooling which relies
on the radiation pressure force. As such, mirror-mediated cooling
uniquely also works in the far-off resonant regime.


\section{Semiclassical Monte-Carlo simulations}

In addition to the perturbative solution of the system dynamics, we
also performed numerical Monte-Carlo simulations. These
semiclassical simulations have a number of advantages: (i) they
allow us to include the harmonic dipole trap consistently, (ii) they
provide solutions to any order in the coupling $g^2/\Delta$ and in
the velocity $v$, and (iii) they include momentum and photon number
diffusion. On the other hand, the numerical treatment requires us to
restrict the analysis to a discrete set of equally spaced modes of
angular frequencies $\omega_k$.

The corresponding set of equations is derived following the approach
of Refs.\ \onlinecite{sde} which we only very briefly outline here.
Starting from the full quantum master equation including quantized
atomic motion and a Liouville-type term for spontaneous atomic
decay, a Wigner transform is applied to obtain a Fokker-Planck
equation for the joint Wigner function of the complex mode field
amplitudes $\alpha_k$ and the atomic momentum $p$ and position $x$.
In the semiclassical approximation, this Fokker-Planck equation is
equivalent to the following set of stochastic differential
equations:
 \bea
 \mathrm{d} x &=& \frac{p}{m} \mathrm{d}t, \label{eq:sde_x}\\
 \mathrm{d} p &=& i\gamma\big[\mathcal{E}(x)\frac{\mathrm{d}}{\mathrm{d} x}\mathcal{E}^*(x)-\mathcal{E}^*(x)\frac{\mathrm{d}}{\mathrm{d} x}\mathcal{E}(x)\big]\mathrm{d}t\nonumber\\
     & &-U_0\big[\mathcal{E}(x)\frac{\mathrm{d}}{\mathrm{d}x}\mathcal{E}^*(x)+\mathcal{E}^*(x)\frac{\mathrm{d}}{\mathrm{d}x}\mathcal{E}(x)\big]\mathrm{d}t\nonumber\\
     & &+k_{t}(x-x_0)\mathrm{d}t+\mathrm{d}P, \label{eq:sde_p}\\
 \mathrm{d}\alpha_k &=& i\Delta_k\alpha_k \mathrm{d}t-(iU_0+\gamma)\mathcal{E}(x)f_k^*(x)\mathrm{d}
        t+\mathrm{d}A_k \label{eq:sde_a}, \quad
 \eea
where $\mathcal{E}(x)=\sum_k \alpha_k \sin(\omega_k x/c)$ is the
total field amplitude at $x$, $x_t$ and $k_t$ are the position of
the trap center and the trap spring constant, respectively, and
$\gamma$ and $U_0$ are the atomic scattering rate and the optical
potential per photon, respectively. The terms $\mathrm{d}P$ and
$\mathrm{d}A_k$ are correlated stochastic white-noise terms
describing the spontaneous redistribution of photons between modes
by the scattering atom and the subsequent fluctuations in atom
momentum and modal photon numbers. They are given by \cite{sde}
 \bea
 \mathrm{d}P & = &  k_0 \sqrt{4\gamma/5} |\mathcal{E}(x)| \mathrm{d}W_0
          +\sqrt{2\gamma} |\frac{\mathrm{d}}{\mathrm{d} x}\mathcal{E}(x)| \mathrm{d}W_+, \\
 \mathrm{d}A_k & = & \sqrt{\gamma/2} \sin\left(\frac{\omega_k x}{c}\right)
            \frac{\frac{\mathrm{d}}{\mathrm{d} x}\mathcal{E}(x)}{|\frac{\mathrm{d}}{\mathrm{d} x}\mathcal{E}(x)|}
            (i\mathrm{d}W_+ - \mathrm{d}W_-), \quad\quad
 \eea
where $\mathrm{d}W_i$ ($i=0,+,-$) are independent stochastic Ito
increments with zero mean, $\langle \mathrm{d}W_i\rangle=0$, and
unit variance, $\langle \mathrm{d}W_i^2\rangle=dt$ \cite{Gardiner2}.

Unfortunately, discretization of the mode frequencies also implies
that the light field is periodic in time with a periodicity given by
the inverse spectral mode spacing $2\pi/\Delta\omega$. It is thus
not possible to follow a single simulation of the stochastic
equations (\ref{eq:sde_x})-(\ref{eq:sde_a}) to a quasi-stationary
state. Instead, we perform averages over short propagation times for
ensembles of different initial temperatures and derive linear
approximations for the rate of temperature change
$\mathrm{d}T/\mathrm{d}t$. Numerically, we found that this approach
works well, however great care is required in the data analysis and
fitting routines as outlined in the following.

\begin{figure}[tb]
  \includegraphics[width=7cm]{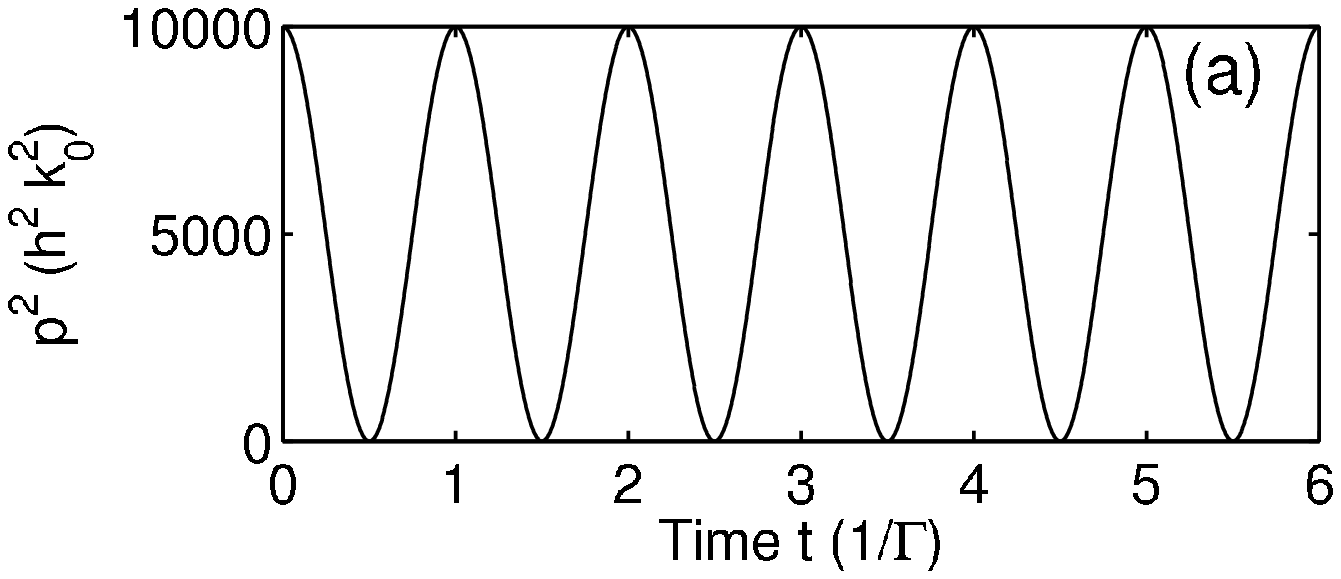}
  \includegraphics[width=7cm]{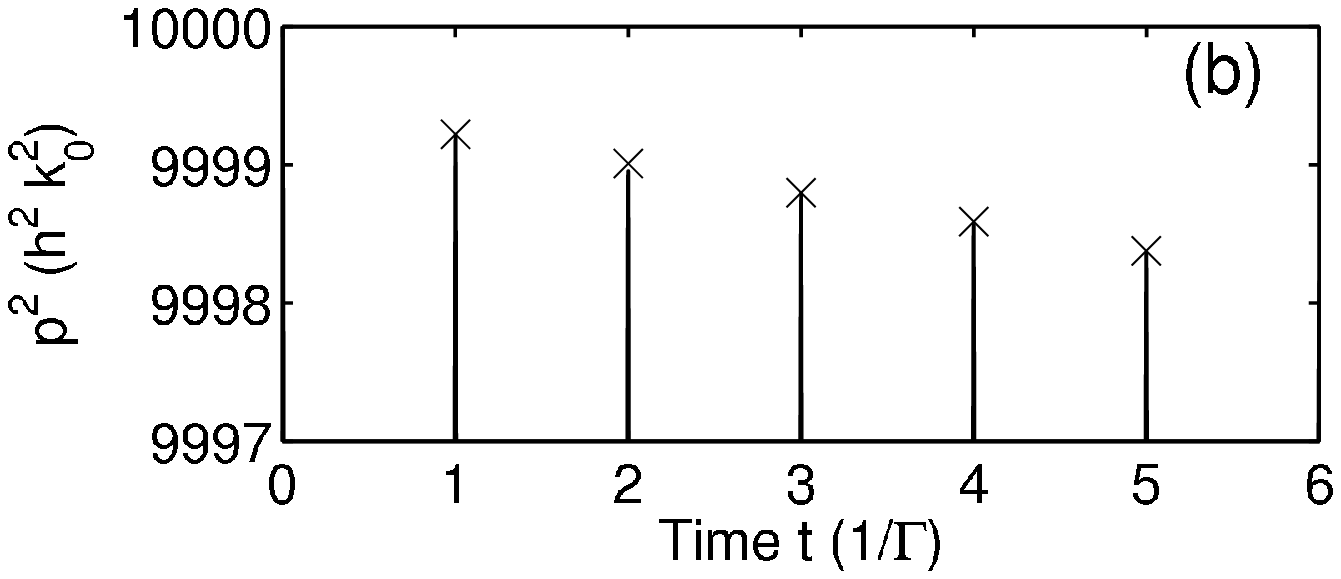}
  \includegraphics[width=7cm]{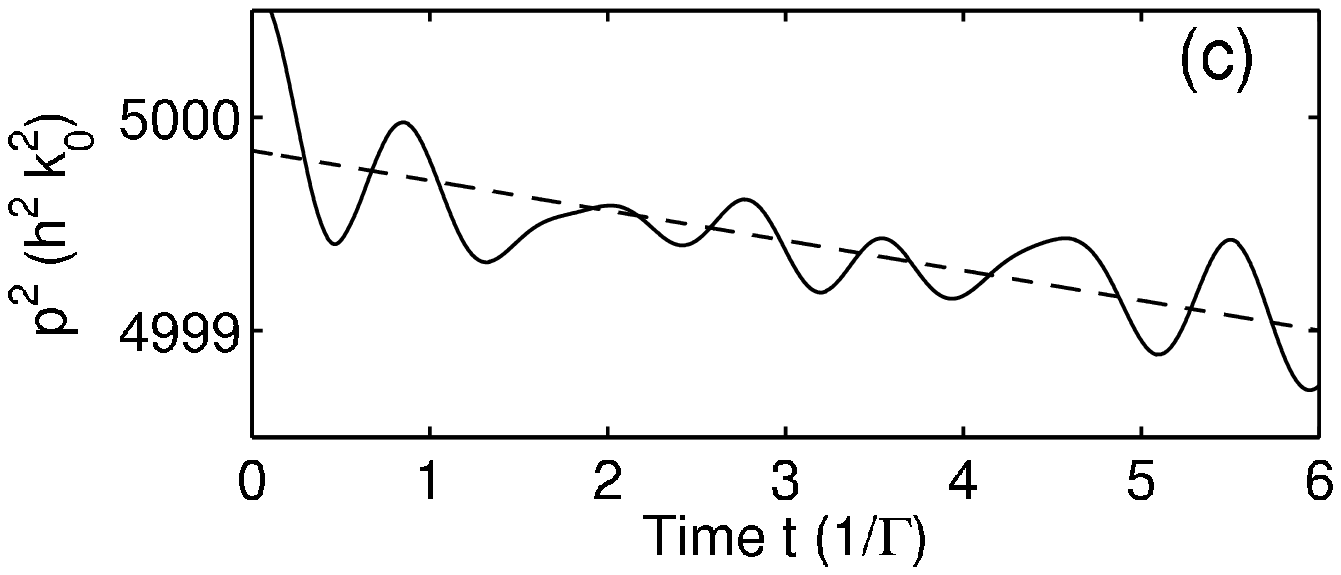}
  \caption{Momentum of a single-particle trajectory in the absence
  of diffusion. (a) Square of momentum versus time. (b) Detail of
  (a) with fitted turning points (crosses) using method 1. (c)
  Fitting method 2: squared momentum after subtraction of a
  least-square fit with a sine function (solid line) and linear fit
  to the result (dashed). Parameters are for Rb atoms,
  $\Delta=-10\Gamma$, $\tau=0.25/\Gamma$, $w=0.7~\mu$m,
  $\Delta\omega=0.1\Gamma$, $\omega_t=0.5\times 2\pi\Gamma$,
  $s=0.076$. The trap is centered at a point of maximum friction.
  }
  \label{fig:fitting}
\end{figure}

Figure \ref{fig:fitting} shows the principles of our data analysis
routines for the example of a single particle trajectory, where
momentum diffusion was neglected for the sake of clarity. An atom
with an initial momentum of 100 $\hbar k_0$ oscillates in a trap
with trap frequency $\omega_t$, Fig.\ \ref{fig:fitting}(a). A closer
look at the maxima of the momentum oscillations, Fig.\
\ref{fig:fitting}(b), reveals the cooling effect due to friction.
Note that this effect is small, only of the order of $10^{-4}$ for
the period of time shown here. It is therefore necessary to remove
the fundamental oscillation with very high accuracy from the
simulated data, as it can otherwise easily mask the effects of
friction.

One possibility (method 1) is to fit each single oscillation with a
harmonic motion. From these fits the positions and momentum
amplitudes of the individual oscillations are obtained, as indicated
by the crosses in the figure. Finally, a linear fit to these data
points provides an accurate measure of the averaged friction
coefficient which can be compared to the analytic result
(\ref{eq:rhoav}). An alternative method (method 2) to extract the
friction coefficient from a trajectory, or from an ensemble average
over many trajectories in the presence of noise, is shown in Fig.\
\ref{fig:fitting}(c). In this case, a \textit{single} sine function
is fitted to the trajectory and subtracted from it. The result is a
curve where the large amplitude oscillations have been removed. A
linear fit to this curve reveals the average slope and thus the
friction coefficient. Both methods, 1 and 2, work well within
certain parameter limits, with method 1 in general being more
accurate but method 2 being quicker to evaluate.

\begin{figure}[tb]
  \includegraphics[width=7cm]{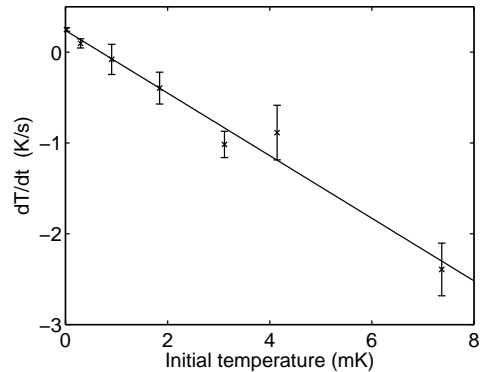}
  \caption{Cooling rate $\mathrm{d}T/\mathrm{d}t$ versus
  initial temperature. The data points (crosses) and error bars are
  obtained from Monte-Carlo simulations using $10^4$ trajectories
  per initial temperature. The solid line is a linear fit to the
  numerical data.  The parameters are as in Fig.\
  \ref{fig:fitting}.
  }
  \label{fig:simu}
\end{figure}

The results of one set of simulations of the full system of
stochastic differential equations (\ref{eq:sde_x})-(\ref{eq:sde_a})
are shown in Fig.\ \ref{fig:simu}. Simulations were performed in
ensembles of $10^4$ independent trajectories, where the initial
conditions of each ensemble were chosen to represent an atomic cloud
at a given temperature. Every trajectory was propagated for a time
60 $\Gamma^{-1}$, and ensemble averages of the squared momentum were
taken as a measure of the ensemble temperature as a function of
time. Finally, method 2 as outlined above was employed to extract a
linear approximation to the cooling rate $\mathrm{d}T/\mathrm{d}t$.
The numerical errors related to the finite number of simulations per
ensemble were estimated by applying the same analysis to
sub-ensembles of $10^3$ trajectories.

Figure \ref{fig:simu} shows that the numerically obtained values of
$\mathrm{d}T/\mathrm{d}t$ are well approximated by a linear function
of initial temperature. This linear behavior is expected for
Brownian motion with linear friction. For very small initial
temperatures, momentum diffusion dominates and the ensemble
temperature increases with time. For large initial temperatures, on
the other hand, friction dominates and the ensemble is cooled. The
stationary temperature where the two effects cancel is obtained as
$T=0.69\pm 0.17$ mK from the linear fit to the data with Gaussian
error propagation. By comparison, the simple analytic estimate
(\ref{eq:Tanal}) predicts a temperature of 0.76 mK, which is in
surprisingly good agreement given the number of approximations made
in the derivation of the perturbative result.


\section{Outlook: Application to Micro- and Nanodevices}

In the analysis presented above, the delay time was provided by a
meter-scale distance between the particle and the mirror. However,
we envisage various routes for integration of this cooling scheme
into microchip devices, as outlined in the following.

\textit{Microresonators.} The necessary delay time can be
conveniently achieved using an integrated optical resonator
\cite{Vahala,Schmiedmayer,Reichel,Mabuchi,Baumberg} instead of a
delay line. For example, a 10 ns delay requires a resonator quality
factor of $Q\approx 10^7$, which is well within the limits of
state-of-the-art microsphere or microdisk resonators. The friction
coefficient $\rho$ will then correspond to the average of Eq.\
(\ref{eq:rhox}) over the delay time $\tau$ for light exiting the
resonator after 1, 2, 3, etc.\ roundtrips.

\textit{Plasmonic field enhancement.} In free space, the minimum
beam diameter is limited by diffraction to about one optical
wavelength. However, it has been shown that microantennas can vastly
enhance local field intensities by plasmon effects, and enhancement
factors of 300 have already been demonstrated \cite{Fischer}. If the
particle could be placed inside such a microantenna, the geometric
factor $\sigma_a/(\pi w^2)$ would effectively be increased by this
factor, leading to significantly faster, more efficient cooling.

\textit{Optomechanics.} The scheme discussed in this paper is
concerned with cooling of microscopic particles. However,
conceptually the proposed cooling method could also be applied to
larger objects, e.g., micromechanical mirrors or oscillators
\cite{Corbitt,Schliesser,Xuereb} by appropriate device design and
scaling of parameters.

With such generalizations we expect that this cooling scheme may be
exploited in integrated atom chips \cite{atomchip}, quantum
information processors \cite{QIP}, atomic clocks \cite{clock}, or
interferometric sensors \cite{sensor}.


\section{Conclusions}

We have analyzed theoretically an optical cooling scheme for neutral
particles which, in principle, is not restricted to atoms but can be
extended to any polarizable species. The scheme exploits the finite
delay-time of light propagating from the particle to a mirror and
back to the particle, and is thus non-Markovian in nature, in
contrast to free-space laser cooling or cavity-mediated cooling
methods. Finally, we proposed extensions of the cooling scheme to
chip-based devices, which will open the road towards practical
applications.\\


\begin{acknowledgments}

The authors acknowledge support by the network on ``Cavity-Mediated
Molecular Cooling'' within the EuroQUAM programme of the European
Science Foundation (ESF) and by the UK Engineering and Physical
Sciences Research Council (EPSRC).
\end{acknowledgments}


\end{document}